\documentclass[twocolumn,showpacs,preprintnumbers,amsmath,amssymb]{revtex4}
\usepackage{tipa}
\usepackage{graphicx}
\usepackage{dcolumn}
\usepackage{bm}

\begin{document}

\title{Interpretation of $D_{sJ}(2632)^+$, $D_{s1}(2700)^\pm$, $D^\star_{sJ}(2860)^+$ and $D_{sJ}(3040)^+$}
\author{Bing Chen$^1$\footnote{chenbing@shu.edu.cn}, Deng-Xia Wang$^1$ and Ailin Zhang$^{1,2}$\footnote{Corresponding author:
zhangal@staff.shu.edu.cn}} \affiliation{$^1$Department of Physics,
Shanghai University, Shanghai 200444, China\\
$^2$Kavli Institute for Theoretical Physics of China, CAS, Beijing
100190, China}

%\date{\today}

\begin{abstract}
$D_s$ mesons are investigated in a semi-classic flux tube model where the spin-orbit interaction is taken into account. Spectrum of D-wave $D_s$ is predicted. The predicted spectrum is much lower than most previous predictions. Analysis of some $D_s$ candidates is made. $D_{sJ}(2632)^+$ may be a $1^-$ ($j^P={3\over 2}^-$ or $1~^3D_1$) orbitally excited $D_s$ meson. $D_{s1}(2700)^\pm$ is very possible the first radially excited $1^-$ ($j^P={1\over 2}^-$ or $2~^3S_1$) $D_s$ meson (the first radial excitation of $D^{\star\pm}_s(2112)^0$). $D^\star_{sJ}(2860)^+$ should be a $3^-$ ($j^P={5\over 2}^-$ or $1~^3D_3$) orbitally excited $D_s$ meson, and $D_{sJ}(3040)^+$ is the first radially excited $1^+$ ($j^P={1\over 2}^+$) $D_s$ meson. Our conclusions are consistent with the most lately BaBar experiment.
\end{abstract}

\maketitle

\section{Introduction}
$D_s$ mesons have arisen people's great interest. More and more $D_s$ mesons were observed by experiments, and many theoretical explorations were triggered. However, some candidates have not been pinned down, their features are not clear yet.

$D^{\star}_{s0}(2317)^\pm$ was first observed by BaBar~\cite{babar1} in $D^\star_{s0}(2317)\to D^+_s\pi^0$ with mass near $2.32~GeV$ in 2003, $D_{s1}(2460)^\pm$ was also first reported by CLEO~\cite{cleo1} in
$D_{s1}(2460)^\pm\to D^\star_s\pi^0$ in 2003. Though there are controversial interpretations of this two states, $D^{\star}_{s0}(2317)^\pm$ and $D_{s1}(2460)^\pm$ are believed to be the $0^+$ and $1^+$ $D_s$ mesons, respectively.

Another surprisingly narrow charmed strange meson, $D_{sJ}(2632)^+$,
was reported by SELEX~\cite{selex} in 2004 in
\begin{eqnarray*}
D^+_{sJ}(2632)\to D^+_s\eta~,~D^0K^+
\end{eqnarray*}
with $M=2632.5\pm 1.7(stat)\pm 5.0(syst)$ and $\Gamma<17$ MeV with
$90\%$ confidence level. This state has an exotic relative branching ratio
$\Gamma(D^0K^+)/\Gamma(D^+_s\eta)=0.14\pm 0.06$. The decay favors the $D_s\eta$ mode over the DK, but the two
channels share the same quark flavors and similar phase space.

Two new $D_s$ mesons were observed in 2006.
$D_{s1}(2700)^\pm$ was first observed by Belle~\cite{belle2} in
$B^+\to \bar D^0D_{s1}\to\bar D^0D^0K^+$ with $M=2715\pm
11^{+11}_{-14}$ and $\Gamma=115\pm 20^{+36}_{-32}$ MeV. The reported
mass and decay width changes a little in the published
paper~\cite{belle3}. $X(2690)$ was reported by
BaBar~\cite{babar3}, but the significance of the signal was not
stated. This state is included in PDG08~\cite{pdg08} with $M=2690\pm
7$ MeV, $J^P=1^-$ and full width $\Gamma=110\pm 27$ MeV.

$D_{sJ}(2860)$ (not listed by the PDG08) was first reported by
BaBar~\cite{babar3} in $D_{sJ}(2860)\to D^0K^+~,~D^+K^0_s$ with
$M=2856.6\pm 1.5(stat)\pm 5.0(syst)$ and $\Gamma=48\pm 7(stat)\pm
10(syst)$ MeV. It was supposed to have natural spin-parity: $J^P=0^+,~1^-,~\cdots$. This state has not been confirmed by Belle, therefore whether it exists is not clear. In fact, there is another possibility for the non-observation of $D_{sJ}(2860)$ by Belle. The non-observation may indicates a high spin for $D_{sJ}(2860)$ that suppresses its
production in B decays.

Very recently, BaBar~\cite{babar4} reports the study of $D_{sJ}$ decays to $D^\star K$ in inclusive $e^+e^-$ interactions. In the report, they observed the decays $D^\star_{s1}(2710)^+\to D^\star K$ and $D^\star_{sJ}(2860)\to D^\star K$. They performed an angular analysis of this two states and measured their branching fractions relative to the $DK$ final state
\begin{equation}\label{br1}
%\begin{split}
\frac{{\cal B}(D^*_{s1}(2710)^+ \to D^*K)}{{\cal B}(D^*_{s1}(2710)^+ \to D K)} =
0.91 \pm  0.13_{stat} \pm  0.12_{syst},
%\end{split}
\end{equation}
\begin{equation}\label{br2}
%\begin{split}
\frac{{\cal B}(D_{sJ}^{*}(2860)^+ \to D^*K)}{{\cal B}(D_{sJ}^{*}(2860)^+ \to D K)} =
1.10 \pm 0.15_{stat} \pm 0.19_{syst}.
%\end{split}
\end{equation}

The new experiment will definitely give more information about this two states. In BaBar experiment, a new broad structure ($D_{sJ}(3040)^+$) at a mass $3044\pm 8(stat)(^{+30}_{-5})(syst)$ MeV with $\Gamma=(239\pm 35(stat)(^{+46}_{-42})(syst))$ is also observed.

So far, it is believed that the S-wave and the P-wave $D_s$ mesons have been established. Resonance beyond the S-wave and the P-wave $D_s$ has not yet been confirmed. How to put these new observed states in the $D_s$ zoo or other family deserves systematic study. In this paper, we study the spectra of $D_s$ mesons and classify these new states in an improved classic flux tube model.

The paper is constructed as follows. In the first section, we give a brief introduction to relevant experiments. $D_s$ mesons are systematically studied in an improved classic flux tube model in the Sec. II. In Sec. III, the new states are classified. Our conclusions and discussions come in the final section.

\section{$D_s$ in classic flux tube model}

The classic flux tube model was studied twenty years ago~\cite{olsson}, the quantization of this model was also performed though the procedure is a little complicated~\cite{olsson1}. Selem and Wilczek studied the light hadrons in this classic flux tube model (mass loaded flux tube model)~\cite{sw}, where the spin-orbit interaction is ignored and the heavy-light hadrons have not investigated. The spin-orbit interaction was taken into account to study the $D$ , $D_s$ and $\Lambda_c$~\cite{zhang1,zhang2}, but the spin-orbit interaction was simplified as a simple $L\cdot S$ coupling. In this paper, the spin-orbit interaction inspired by QCD will be employed.

In a relativized quark model, the quark-antiquark potential $V(\vec{r})$ inside mesons with one heavy quark was written as~\cite{gk}:
\begin{eqnarray*}
V(\vec{r})=H^{Conf}_{q\bar{q}}+H^{Cont}_{q\bar{q}}+H^{ten}_{q\bar{q}}+H^{SO}_{q\bar{q}},
\end{eqnarray*}
where the spin-orbit interaction is
\begin{eqnarray*}
H^{SO}_{q\bar{q}}=H^{SO+}_{q\bar{q}}(\vec{S}_q+\vec{S}_{\bar{q}})\cdot
\vec{L}+H^{SO-}_{q\bar{q}}(\vec{S}_q-\vec{S}_{\bar{q}})\cdot
\vec{L}.
\end{eqnarray*}
For mesons
\begin{eqnarray*}
H^{SO+}_{q\bar{q}}=[\frac{2\alpha_s}{3r^{3}}(\frac{1}{m_q}+\frac{1}{m_{\bar{q}}})^{2}-\frac{1}{4r}\frac{\partial
H^{conf}_{q\bar{q}}}{\partial
r}(\frac{1}{m^{2}_{q}}+\frac{1}{m^{2}_{\bar{q}}})],
\end{eqnarray*}
and
\begin{eqnarray*}\label{eq2}
H^{SO-}_{q\bar{q}}=(\frac{2\alpha_s}{3r^{3}}-\frac{1}{4r}\frac{\partial
H^{conf}_{q\bar{q}}}{\partial
r})(\frac{1}{m^{2}_{q}}-\frac{1}{m^{2}_{\bar{q}}}).
\end{eqnarray*}

The mass of mesons is obtained from the $Schr\ddot{o}dinger$ equation where the eigenstates of $(J^2,L^2,S^2,J_z)$ is employed. There is no mixing for $^3P_0$ and the $^3P_2$, but there is a mixing between the $^3P_1$ and the $^1P_1$. This two mixed states are denoted as $(^3P_1)'$ and $(^1P_1)'$, respectively. The case is similar for the D-wave and F-wave mesons. In the limit where $m_Q\to\infty$, we obtain the mass formulas
\begin{eqnarray}
 \left(
           \begin{array}{cc}
                    M_1-H^+  & {\sqrt{2}\over 3}H^- \\
                    {\sqrt{2}\over 3}H^-  & M_1 \\
                    \end{array}
     \right)  \left(
           \begin{array}{c}
                     ^3P_1 \\
                     ^1P_1 \\
                    \end{array}
     \right)= \left(
           \begin{array}{c}
                     M(^3P_1)^\prime \\
                     M(^1P_1)^\prime \\
                    \end{array}
     \right)
\end{eqnarray}
for the P-wave multiplet, and
\begin{eqnarray}
  \left(
           \begin{array}{cc}
                    M_2-H^+  & \sqrt{\frac{2}{3}}H^- \\
                    \sqrt{\frac{2}{3}}H^-  & M_2 \\
                    \end{array}
     \right)  \left(
           \begin{array}{c}
                     ^3D_2 \\
                     ^1D_2 \\
                    \end{array}
     \right)= \left(
           \begin{array}{c}
                     M(^3D_2)^\prime \\
                     M(^1D_2)^\prime \\
                    \end{array}
     \right)
\end{eqnarray}
for the D-wave multiplet. In these equations, $H^\pm$ ($=\langle H^{SO+}\rangle$) is the expectation value of the spacial part of $H^{SO\pm}$ and $M_L~(L=1,2)$ is the center of mass of the multiplet which is independent of the spin-orbit interaction. In the framework of $L\cdot S$ coupling scheme, the calculable $<H^{SO+}>\approx <H^{SO-}>$ (denoted as $a$) when the heavy quark effect is considered. The difference of $a$ among different orbits could be ignored in this case.

Accordingly, mass of all the mesons could be written as
\begin{eqnarray}\label{eq1}
M(^{2S+1}L_J)=M_L+\xi_{L,S}a,
\end{eqnarray}
where $M(^{2S+1}L_J)$ is the mass of the physical state and $\xi_{L,S}$ is the calculable coefficient.

In Table.I, the explicit $\xi_{L,S}$ for P-wave and D-wave mesons is
estimated. The states with an upper prime correspond to the
mixing physical states.
\begin{table}
\begin{tabular}{llllllllllllllll}
\hline\hline state& & & & &$\xi_{L,S}$& & & & &state& & & & &$\xi_{L,S}$ \\
\hline\hline
$(^3P_2)$& & & & &$+1.00$& & & & &$(^3D_3)$& & & & &$+2.00$ \\
$(^3P_1)'$& & & & &$+0.20$& & & & &$(^3D_2)'$& & & & &$+0.46$ \\
$(^1P_1)'$& & & & &$-1.20$& & & & &$(^1D_2)'$& & & & &$-1.46$ \\
$(^3P_0)$& & & & &$-2.00$& & & & &$(^3D_1)$& & & & &$-3.00$ \\
\hline\hline
\end{tabular}
\caption{$\xi_{L,S}$ for P-wave and D-wave mesons} \label{table-1}
\end{table}

When Eq.~(\ref{eq1}) is compared with Eq.~(3) in Ref.~\cite{zhang1} (or Eq.~(2) in Ref.~\cite{zhang2}), it is reasonable to improve the formula of mass of $D_s$ meson in the classic flux tube model to
\begin{eqnarray}\label{eq2}
E=M_c+\sqrt{\frac{\sigma L}{2}}+2^{\frac{1}{4}}\kappa
L^{-\frac{1}{4}}m^{\frac{3}{2}}_s+\xi_{L,S}a,
\end{eqnarray}
where $M_c$ is the $c$ quark mass, and parameter $a$ could be fixed through experimental data. Obviously, the first two terms at the right side of Eq.~(\ref{eq2}) is the center of mass of the multiplet resulting from confinement, and the last term at the right side of Eq.~(\ref{eq2}) results from the spin-orbit contribution.

The next step is to fix the parameters in Eq.~(\ref{eq2}). For this purpose, some states are used as inputs. As well known, scalar meson usually has special features, so the $0^+$ $D^\star_{s0}(2317)^\pm$ is not used as an input. In the fitting procedure, $D_{s1}(2536)^\pm$, $D_{s1}(2460)^\pm$, $D_{s2}(2573)^\pm$ and $D^+_{sJ}(2860)$ are identified as the four states $(^3P_1)^\prime$, $(^1P_1)^\prime$, $^3P_2$ and $^3D_3$, respectively. The reason for the identification of $D^+_{sJ}(2860)$ is stated in the next section.

Firstly, the $a$ in Eq.~(\ref{eq2}) is fixed: $a=0.05$ GeV, through the mean square of
\begin{eqnarray*}
\Delta(M(^3P_2)-M(^3P_1)^\prime)=0.04~GeV,~a=0.05~GeV;\\
\Delta(M(^3P_2)-M(^1P_1)^\prime)=0.11~GeV,~a=0.05~GeV;\\
\Delta(M(^3P_1)^\prime-M(^1P_1)^\prime)=0.08~GeV,~a=0.06~GeV.
\end{eqnarray*}

Other three parameters $M_c$, $m_s$ and $\sigma$ can not be fixed through the four states (only P-wave and D-wave). Since $\sigma$ reveals the dynamics of confinement of meson, it is reasonable to borrow its value from other quark model ($\sigma$ varies little in different model). Here, $\sigma\approx 1.10$ GeV$^2$~\cite{gi,zhang1} is employed. The $M_c$ and $m_s$ is therefore determined as: $M_c=1.40~GeV$ and $m_s=0.42~GeV$. The parameters $M_c$, $m_s$ and $a$ are comparable with Refs.~\cite{zhang1,zhang2}.

With these parameters in hand, masses of other D-wave $D_s$ could be predicted. All the candidates of S, P and D-wave $D_s$ and their spectra are shown in Table.II. In the table, the predictions in Refs.~\cite{gi,pe} are listed. The spectra of D-wave $D_s$ obtained here is much lower than most theoretical predictions. Similar lower spectra was obtained in a relativistic chiral quark model ten years ago~\cite{gr}. In a popular viewpoint, the origin of the lower mass of $D^\star_{s0}(2317)^\pm$ and $D_{s1}(2460)^\pm$ is that coupled-channel effects can shift masses from naive quark model predictions by up to a couple hundred MeV. From our analysis, it is found that there is another possibility. New feature of the confinement potential in heavy-light system may result in a lower spectra.

In fact, it is very possible that $D_{s1}(2460)^\pm$ and $D_{s1}(2536)^\pm$ are not the $j^P={1\over 2}^+$ $1^+$ and the $j^P={3\over 2}^+$ $1^+$ $D_s$, respectively. The $j^P={1\over 2}^+$ $1^+$ $D_s$ was predicted to have a large width in Ref.~\cite{pe}, while the full width is very small ($<3.5$ MeV) in PDG08~\cite{pdg08}. $D_{s1}(2460)^\pm$ and $D_{s1}(2536)^\pm$ are also mixing states of the $j^P={1\over 2}^+$ $1^+$ and the $j^P={3\over 2}^+$ $1^+$ $D_s$. Accordingly, , parenthesis is added for the $j^P$ in Table.II.

\begin{table}
\begin{tabular}{lllllll}
\hline\hline
 Candidates~\cite{pdg08} & $J^P$ & $j^P$ &  $n^{2S+1}L_J$ & GI~\cite{gi} & PE~\cite{pe}
& our paper\\
\hline\hline $D^\pm_s(1969)$ & $0^-$ & ${1\over 2}^-$ & $1^1S_0$ & 1.98 & 1.965 & - \\
$D^{\star\pm}_s(2112)^0$ & $1^-$ & ${1\over 2}^-$ & $1^3S_1$ & 2.13 & 2.113 & -\\
\hline\hline $D^\star_{s0}(2317)^\pm$ & $0^+$ & ${1\over 2}^+$ & $1^3P_0$ & 2.48 & 2.487 & 2.42\\
$D_{s1}(2536)^\pm$ & $1^+$ & $({3\over 2}^+)$ & $(1^3P_1)^\prime$ & 2.57 & 2.535 & 2.53\\
$D_{s1}(2460)^\pm$ & $1^+$ & $({1\over 2}^+)$ & $(1^1P_1)^\prime$ & 2.53 & 2.605 & 2.46\\
$D_{s2}(2573)^\pm$ & $2^+$ & ${3\over 2}^+$ & $1^3P_2$ & 2.59 & 2.581 & 2.57\\
\hline\hline $D_{sJ}(2632)$ & $1^-$ & ${3\over 2}^-$ & $1^3D_1$ & 2.90 & 2.900 & 2.62\\
? & $2^-$ & $({5\over 2}^-)$ & $(1^3D_2)^\prime$ & - & 2.913 & 2.81\\
? & $2^-$ & $({3\over 2}^-)$ & $(1^1D_2)^\prime$ & - & 2.925 & 2.70\\
$D_{sJ}(2860)$ & $3^-$ & ${5\over 2}^-$ & $1^3D_3$ & 2.92 & 2.953 & 2.87\\
\hline\hline
\end{tabular}
\caption{Spectrum of charmed strange mesons (GeV) with parameters
$\sigma=1.10$ GeV$^2$, $M_c=1.40$ GeV, $m_s=0.42$ GeV and $a=0.05$
GeV.} \label{table-2}
\end{table}

\section{$D_{sJ}(2632)^+$, $D_{s1}(2700)^\pm$, $D^\star_{sJ}(2860)^+$ and $D_{sJ}(3040)^+$}

\subsection{$D_{sJ}(2632)^+$}

$D_{sJ}(2632)^+$ was suggested to be a four-quark state in Ref.~\cite{maiani,li}, it
was interpreted as a conventional $1^-(2^3S_1)$ $c\bar s$ in Ref.~\cite{zhao,swanson,zhong}. It was pointed out in Ref~\cite{zhang}, $D_{sJ}(2632)^+$ seems unlike the $1^-(2^3S_1)$ $c\bar s$. However, $D_{sJ}(2632)^+$ is not observed by BaBar~\cite{babar5}, FOCUS or Belle, it seems that this state is excluded. In the semi-classic flux tube model, the $1^-$ (${3\over 2}^-$ or $1^3D_1$) $D_s$ with mass $2.62$ GeV is predicted. In a framework of $^3P_0$ pair creation model~\cite{ki}, the decay $D_{s1}(1^3D_1)\to D^0K$ of the predicted state is found to have width $\Gamma(D_{s1}(1^3D_1)\to D^0K)=3.73$ MeV. It is very possible that $D_{sJ}(2632)^+$ is the $1^-$ $1^3D_1$.

\subsection{$D_{s1}(2700)^\pm$}

There are three possible sets of experimental data
for $D_{s1}(2700)^\pm$ as enumerated in the introduction, it is
reasonable to regard them as the same state because they have the
approximately equal mass and decay width. In Ref.~\cite{zhu}, $D_{s1}(2700)^\pm$ was thought probably the $1^-(1~^3D_1)$ $D_s$, which is $\approx 200$ MeV lower than theoretical predictions~\cite{gi,pe}. $D_{s1}(2700)^\pm$ was interpreted as the $1^-(2~^3S_1)$ $D_s$~\cite{ctls,zhang3} (first radial excitation of the $D^\star_s(2112)^\pm$). When the observed branching ratio (Eq.~(\ref{br1})) is compared with theoretical predictions~\cite{cfnr}, the $1^-(2~^3S_1)$ assignment is preferred.

\subsection{$D^\star_{sJ}(2860)^+$}

$D^\star_{sJ}(2860)^+$ was once interpreted as a conventional $0^+~(2^3P_0)$ $c\bar s$~\cite{br,zhu}, which is also $\approx 200$ MeV lower than the theoretical prediction~\cite{pe}. This state was interpreted as a conventional $3^-~(1^3D_3)$ $c\bar s$~\cite{cfn,zhu,zhang3}. The observation of $D^\star_{sJ}(2860)^+\to D^\star K$~\cite{babar4} rules out the possibility of $0^+~(2~^3P_0)$. In the meantime, the observed branching ratio (Eq.~(\ref{br2})) is in significant disagreement with theoretical predictions~\cite{ctls}. On the other hand, $\frac{{\cal B}(D_{sJ}^\star (2860)^+ \to D^\star K)}{{\cal B}(D_{sJ}^\star(2860)^+ \to D K)} =1.23$
was obtained when $D^\star_{sJ}(2860)^+$ is treated as a
$2~^3S_1$~\cite{cfn}, which is in agreement with (Eq.~(\ref{br2})). However,
one expects this vector meson to have a considerably lower mass,
around $2720$ MeV~\cite{gi,tala}, which makes the $D_{s1}(2700)^\pm$ a
much better candidate. Due to these facts, the existence of two largely
overlapping resonances at about $2.86$ GeV (radially excited tensor $2^+$ and radially excited scalar $0^+$ $c\bar s$ states) was suggested~\cite{br2}. It was argued that the possibility of $0^+~(2~^3P_0)$ could not be ruled out now. In our analysis, $D^\star_{sJ}(2860)^+$ is an excellent candidate for the $3^-~(1^3D_3)$ $c\bar s$.

\subsection{$D_{sJ}(3040)^+$}

Both the nonobservation of $D_{sJ}(3040)^+\to DK$ and the angular analysis suggest an unnatural parity $J^P=0^-,~1^+,~2^-,~\cdots$ for $D_{sJ}(3040)^+$. In Ref.~\cite{pe}, the first radial excitation of the  $1^+~(j^P={1\over 2}^+)$ $D_s$ ($\approx 3.165$ GeV) was predicted to have a large width $\Gamma\approx 210$ MeV, the first radial excitation of the  $1^+~(j^P={3\over 2}^+)$ $D_s$ ($\approx 3.114$ GeV) was predicted to have width $\Gamma\approx 51$ MeV, and the F-wave $D_s$ have not large width. In Ref.~\cite{ctls}, the mass of $D_s(2^3P_1)^\prime$ was predicted around $2995$ MeV. Therefore, it is reasonable to interpret $D_{sJ}(3040)^+$ as the first radially excited  $1^+$ ($j^P={1\over 2}^+$) $D_s$ (or the mixng of the radially excited $1^+$ ($j^P={1\over 2}^+$) and $1^+$ ($j^P={3\over 2}^+$), but mainly ($j^P={1\over 2}^+$)).

In fact, $D^\star_s(2112)^\pm$, $D_{s1}(2460)^\pm$ (or $D_{s1}(2536)^\pm$), $D_{s1}(2700)^\pm$ and $D_{sJ}(3040)^+$ meet trajectories on $(n, M^2)$-plot:
$M^2=M^2_0+(n-1)\mu^2$~\cite{aas,zhang3},
\begin{center}
$M^2((D_{s1}(2700)^\pm)-(D^\star_{s1}(2112)^\pm)=2.78~GeV^2$;
$M^2((D_{sJ}(3040)^+)-(D_{s1}(2460)^\pm))=3.19~GeV^2$;
$M^2((D_{sJ}(3040)^+)-(D_{s1}(2536)^\pm))=2.79~GeV^2$;
\end{center}

Some partners of these observed states are expected to exist. Two $2^-$ D-wave $D_s$ may exist at $2.70$ GeV and $2.81$ GeV. The radially excited $1^+$ and $2^+$ $D_s$ are expected to exist around $3.0$ GeV.

\section{Conclusions and discussions}

$D_s$ mesons are investigated in a semi-classic flux tube model. In this model, the classic flux tube is responsible for the confinement of quark-antiquark inside mesons. The spin-orbit interaction is involved through a deep investigation in the relativized quark model. A formula of the energy for heavy-light mesons is obtained. In terms of the formula, the spectra of D-wave $D_s$ are predicted after some observed states are taken as inputs. The predicted spectra of D-wave $D_s$ are much lower than most previous predictions.

In a popular viewpoint, it is believed that the origin of the lower mass of $D^\star_{s0}(2317)^\pm$ and $D_{s1}(2460)^\pm$ is that coupled-channel effects can shift their masses from naive quark model predictions by up to a couple hundred MeV. Our investigation indicates that the lower spectra of D-wave $D_s$ may implies another possibility. There is new feature for the quark-antiquark potential in heavy-light system, which may result in a lower spectra.

The unidentified $D_{sJ}(2632)^+$, $D_{s1}(2700)^\pm$, $D^\star_{sJ}(2860)^+$ and $D_{sJ}(3040)^+$ are analyzed, possible assignments to them are made. Our conclusions are consistent with the most lately BaBar experiment.

$D_{sJ}(2632)^+$ is very likely the $1^-$ ($j^P={3\over 2}^-$ or $1^3D_1$) orbitally excited $D_s$ meson with a narrow width.

$D_{s1}(2700)^\pm$ may be the $1^-$ ($j^P={1\over 2}^-$ or $2~^3S_1$) $D_s$ meson (the first radially excitation of $D^{\star\pm}_s(2112)^0$).

$D^\star_{sJ}(2860)^+$ is an excellent candidate for the $3^-$ ($j^P={5\over 2}^-$ or $1^3D_3$) $c\bar s$.

$D_{sJ}(3040)^+$ is interpreted as the first radially excited $1^+$ ($j^P={1\over 2}^+$) $D_s$ meson.

Two $2^-$ D-wave $D_s$ at $2.70$ GeV and $2.81$ GeV, and two radially excited $D_s$ around $3.0$ GeV are predicted.

However, $D^+_{sJ}(2860)$ used as an input has not been definitely confirmed by other experimental group. If $D^+_{sJ}(2860)$ does not exist or it is not a $3^-~(1^3D_3)$ $c\bar s$, our prediction of the spectra of the D-wave may change a little. Therefore, more experiments are required to pin down this state.

If the predicted lower spectra is confirmed by experiments in the future, most previous predictions of the spectra deserve re-examination. Lower spectra of the D-wave $D_s$ indicates that there exist unclear features of quark-antiquark potential.

Acknowledgment: This work is supported by the National Natural
Science Foundation of China under the grant: 10775093.

\end{document}